\documentclass[11pt]{aastex}
\usepackage{graphicx,emulateapj5,apjfonts}
\usepackage{onecolfloat}
\usepackage{epsf}

\shortauthors{Bell et al.}
\shorttitle{The Galaxy Baryonic Mass Function}

\newcommand{\hi}{{\rm H{\sc i} }}
\newcommand{\mol}{{\rm H$_2$ }}

\newcommand{\peg}{{\sc P\'egase }}

\slugcomment{{\sc The Astrophysical Journal Letters, Accepted: } January 30, 2003 }

\begin{document}

%%%%% Added the \def\head{ lark.

\def\head{

\title{A First Estimate of the Baryonic Mass Function of Galaxies }

\author{Eric F.\ Bell}
\affil{Max-Planck-Institut f\"ur Astronomie,
K\"onigstuhl 17, D-69117 Heidelberg, Germany; \texttt{bell@mpia.de}}
\author{Daniel H.\ McIntosh, Neal Katz, and Martin D.\ Weinberg}
\affil{Astronomy Department, University of Massachusetts, Amherst, MA 
01003, USA; \texttt{dmac@hamerkop.astro.umass.edu, 
nsk@kaka.astro.umass.edu, weinberg@astro.umass.edu}}

\begin{abstract}
We estimate the baryonic (stellar$+$cold gas) mass function of
galaxies in the local Universe by assigning 
a complete sample of {\it Two Micron All
Sky Survey} and {\it Sloan Digital Sky Survey} galaxies 
a gas fraction based on a statistical sample of the 
entire population, under the assumption of a universally-applicable
stellar initial mass function.
The baryonic mass function is similar to the 
stellar mass function at the high mass end, and has a reasonably
steep faint-end slope owing to the typically high cold
gas fractions and low stellar mass-to-light ratios characteristic
of low-mass galaxies.  The Schechter Function fit parameters
are $\phi^*\,h^{-3} = 0.0108(6)$\,Mpc$^{-3}$\,log$_{10}M^{-1}$, 
$M^*\,h^2 = 5.3(3)\times10^{10}\,M_{\sun}$, and $\alpha = -1.21(5)$, with
formal error estimates given in parentheses.
We show that the \hi and \mol mass functions derived using this indirect
route are in agreement with direct estimates, validating our indirect
method.  Integrating under the 
baryonic mass function and incorporating all sources
of uncertainty, we find that the 
baryonic (stellar$+$cold gas) mass density implied by this
estimate is $\Omega_{\rm cold\,\,baryon} = 
2.4^{+ 0.7}_{- 1.4}\times10^{-3}$, or 8$^{+4}_{-5}$\% of 
the Big Bang nucleosynthesis
expectation.  
\end{abstract}

\keywords{galaxies: general ---  
galaxies: luminosity function, 
mass function --- galaxies: stellar content }
}%%%end head

\twocolumn[\head]

\section{Introduction}

The distribution of the mass in collapsed baryons (cold gas
and stars) in galaxies is a fundamental prediction of galaxy formation  
models.  Unfortunately, to date there is 
no robust estimate of the baryonic mass function (MF) of galaxies,
leaving modelers with the non-trivial task of 
predicting stellar masses or, even worse, galaxy luminosities.
Discrepancy between the model and data may indicate a problem with the
predicted distribution of galaxy baryonic masses, or could represent
poorly-constrained
star formation (SF), stellar population or dust prescriptions.
In this {\it Letter}, we present a first estimate of the
baryonic MF of galaxies by assigning galaxies gas fractions 
statistically (based on an independent sample), under the assumption of  
a universally-applicable stellar initial mass function 
(IMF).\footnote{A time-varying IMF, as 
speculated on by \citet{hernandez01} or \citet{ferguson02},
would invalidate this assumption.}

The time is ripe to attempt this for the first time.  
With the advent of large, relatively complete surveys, the luminosity 
function (LF) 
is now well-constrained in the optical and near-infrared (NIR)
\citep{gardner97,cole01,kochanek01,norberg02,liske03,blanton03,bell03}.
Furthermore, under the assumption of a universally-applicable stellar
IMF, the distribution of stellar masses is reasonably well-constrained,
with an overall normalization uncertainty caused by
our relatively poor knowledge of the faint end slope of the IMF
\citep{cole01,bell03}.  Crucially, there are also relatively
large samples of galaxies with $K$-band data and gas masses,
allowing a reasonably accurate characterization of the gas mass
of galaxies as a function of their physical parameters \citep{bdj}.

\section{The Data and Methodology} \label{sec:data}

Because of the lack of a large galaxy survey with both gas mass
and $K$-band data, we take a sampling
approach, analogous to that used by \citet{loveday}
to estimate the $K$-band luminosity function from a $B$-band
limited survey.  Essentially, 
we estimate a stellar MF (\S \ref{sec:sm}) and then add
representative gas masses to each galaxy (\S \ref{sec:glue}),
allowing us to estimate the distribution of galaxy baryonic masses
(\S \ref{sec:results}).  We assume 
$\Omega_{\rm m} = 0.3$, $\Omega_{\Lambda} = 0.7$, and 
$H_0 = 100 h $\,km\,s$^{-1}$\,Mpc$^{-1}$.
 
\subsection{Estimating the Stellar Mass Function} \label{sec:sm}

We construct the baryonic MF
using a combined sample of galaxies from the 
{\it Two Micron All Sky Survey} \citep[2MASS;][]{skrut} and the
{\it Sloan Digital Sky Survey} \citep[SDSS;][]{york}. 
We use the SDSS early data release \citep[EDR;][]{edr} to 
provide an 84\% redshift complete $r \le 17.5$ sample
of galaxies with accurate $ugriz$ fluxes over 414 square degrees, which is
$\sim$10\% less than the whole EDR imaging area because some
spectroscopic plates that were not attempted \citep{edr}.
The 84\% redshift completeness within this area is our own direct
estimate based on the fraction of galaxies fulfilling 
the \citet{strauss02} criteria that have spectra, 
in agreement with the EDR analysis of \citet{nakamura03}. 
To account for light missed by the Petrosian magnitude estimator
\citep{strauss02,blanton03}, 
we add 15\% to the fluxes of galaxies morphologically classified as
early-type using the SDSS $r$-band
concentration parameter following \citet{strat}.
This correction produces only a $\la 5$\% effect 
on the LFs and MFs \citep{blanton03,bell03}.
We also correct for an $\sim 8$\% overdensity of galaxies in 
the EDR, as estimated by comparing the number of $10 \le K \le 13.5$ galaxies
in the EDR spectroscopic area with that from the sky with $|b| \ge 30\deg$.

We use the now complete 2MASS extended source
and point source catalogs
to augment the SDSS $ugriz$ fluxes with 
$K$-band fluxes, and for extended sources
$K$-band half-light radii.  We correct 2MASS $K$-band fluxes
to total following a comparison 
with deeper $K$-band data from \citet{loveday}; for extended sources
this amounts to a 0.1 mag correction \citep{bell03}.
We do not use 2MASS 
$J$ or $H$-band data here because we cannot correct the magnitudes similarly.
The optical and NIR magnitude zero points 
are accurate to $\sim$0.05 and $\sim$0.02 mag respectively, 
and the random errors are
0.05 mag (optical) and 0.2 mag (NIR).

To estimate
$k$-corrections, evolution corrections, and the pre\-sent
day stellar mass-to-light ratios (M/Ls), we fit the 
$ugrizK$ observed fluxes\footnote{Not all galaxies have 
$ugrizK$ fluxes.  
We have checked that missing passbands do not significantly
bias the estimated $k$-corrections, evolution corrections 
or stellar M/Ls (but do, of course, increase the random error 
somewhat).} with model stellar populations.
These populations have
a range of metallicities and SF histories at a given redshift.  We
use the \peg stellar population synthesis model \citep[see][for 
a description of an earlier version of their model]{fioc97} 
with a `diet' Salpeter IMF \citep[following][]{ml}
that has the same colors and luminosity as a normal
Salpeter IMF, but with only 70\% of the mass (due to a smaller number of 
low-mass stars).
Corrections derived using this technique are consistent
with those used by \citet{blanton03}.
The stellar M/Ls we derive are within 10\% of
those from the spectral modeling technique of \citet{kauffmann03a},
accounting for differences in IMF; the random and systematic
uncertainties from dust and bursts of SF dominate, however, and
are $\la 25$\% \citep{ml}.
This IMF is `maximum disk', inasmuch
as IMFs richer in faint low-mass stars over-predict the rotation velocity
of Ursa Major Cluster galaxies with $K$-band photometry and
well-resolved \hi rotation curves.  This prescription 
thus gives the maximum possible stellar M/L.  Naturally,
a different choice of IMF allows lower M/Ls.  For example, 
the popular Kennicutt or Kroupa IMFs have $\sim$37\% lower M/Ls than
this IMF, and are thus `submaximal' \citep[see][for more discussion 
of this point]{ml}.

We calculate LFs and stellar MFs using the $V/V_{\rm max}$ formalism
\citep{felten}, taking into account foreground Galactic extinction, 
$k$-corrections, and evolution corrections.  In \citet{bell03}, 
we match precisely published LFs; 
in particular, we reproduce the $g$-band and $K$-band LF and luminosity 
densities to within $\la 10$\% \citep{blanton03,cole01,kochanek01}.  
Furthermore, this method produces accurate stellar MFs
that match the estimate of \citet{cole01} to 
$\sim$5\% in total stellar mass density (accounting for 
IMF differences), {\it but can do so using LFs limited by optical 
or NIR magnitude limits} \citep{bell03}.
For this {\it Letter}, we choose 11848 galaxies with 
$13 \le r \le 17.5$ and $g \le 17.74$, which ensures that we have accurate 
$g-r$ color estimates providing a 
stellar M/L accuracy of better than 25\%, while avoiding
potential biases against low surface brightness galaxies in 2MASS 
\citep{bell03}.  
The stellar MF estimated using this technique is shown 
in Fig.\ \ref{fig:mf}, along with the stellar MF 
of \citet{cole01} for comparison.  A much more extensive
description of the LF and stellar MF construction is given by 
\citet{bell03}.

\subsection{Estimating Gas Masses} \label{sec:glue}

Because there are no samples of galaxies with 
good number statistics, deep optical/NIR data
and gas masses, we estimate the gas masses of
SDSS$+$2MASS galaxies indirectly.
We use galaxies from 
\citet{bdj} with $K$-band luminosities, half-light radii
and gas masses to statistically assign a gas mass to every 
SDSS$+$2MASS galaxy, appropriate to its $K$-band luminosity
and half-light radius.  

\begin{figure}[t]
\vspace{-0.5cm}
\hspace{-0.5cm}
\epsfbox{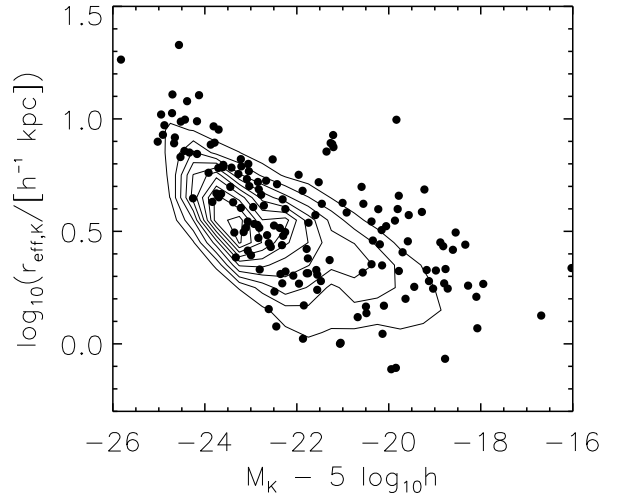}
\vspace{-0.2cm}
\caption{\label{fig:seln} $K$-band magnitude and
half-light radius for the 6573 late-type SDSS$+$2MASS galaxies in our
sample (contours) and 
for the 156 galaxy comparison sample of \citet[filled circles]{bdj}.  
  } \vspace{-0.2cm}
\end{figure}

\begin{figure*}[t]
\hspace{-0.5cm}
\epsfbox{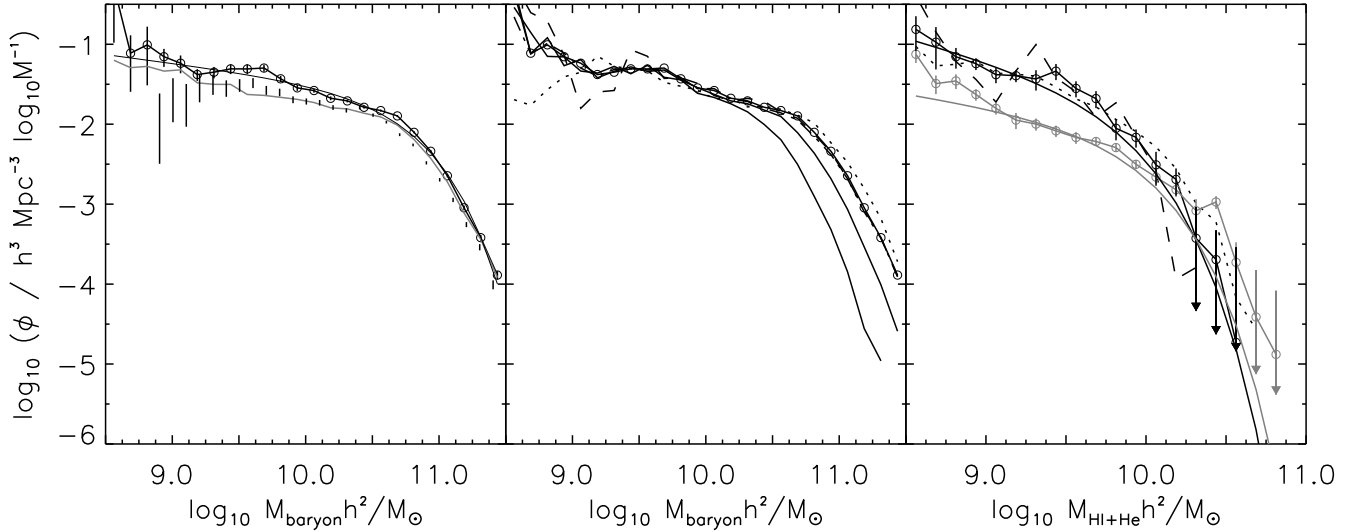}
\vspace{-0.2cm}
\caption{\label{fig:mf}  The baryonic mass function of galaxies.
In the left hand panel, we show the stellar MF of 
\citet[naked error bars]{cole01} corrected to our `maximum disk'
IMF, the stellar MF 
derived using our SDSS $g$-band selected sample (solid grey line),
and the baryonic MF of galaxies, assuming the 
`maximum disk' IMF (solid line with open circles and 
error bars, with a Schechter Fit
as the thin solid line).  
In the middle panel, we show different versions of  the 
baryonic MF, illustrating the different sources of 
uncertainty.  The solid line with open circles again is the baryonic MF
defined using the default gas mass estimation technique,
the dashed line shows the 
effect of choosing the nearest galaxy on the half-light radius--luminosity
plane for estimating the gas mass, the dotted line 
shows the effect of allowing SDSS$+$2MASS early-type galaxies to also have gas,
and the two solid lines are the effect of choosing a
Kennicutt or Kroupa IMF (higher line), or a M/L consistent
with a Bottema disk (lowest line). 
In the right hand panel we test the gas mass estimation 
method by comparing the \hi and \mol MFs predicted using 
this method with observations.  
The solid line with open circles and error bars
denotes the match of SDSS$+$2MASS spirals with 
randomly-selected nearby galaxies, and the dotted and dashed lines
are as in the middle panel.  The solid black curve
is the blind \hi MF of \citet{rosenberg}.
The \mol galaxy MF is shown as the grey line with open circles
and error bars, and the Schechter Fit to the first observational
estimate of the \mol MF \citep[from][]{keres03} is shown as the 
solid grey line. } \vspace{-0.4cm}
\end{figure*}

Fig.\ \ref{fig:seln} shows
the $K$-band half-light radii and luminosities
for the late-type subsample of the 
SDSS$+$2MASS galaxies (contours) and for the comparison
sample of 156 galaxies with gas masses (filled circles) taken from \citet{bdj}.
We estimate gas masses by multiplying the \hi gas mass by 
1.33 to account for helium, and by the morphological type-dependent
\mol to \hi ratio presented by \citet{young89}.  
For the 21\% of the galaxies without $K$-band luminosities 
and sizes, we adopt the $r$-band half-light
radii and estimate the $K$-band luminosity by dividing the 
$g$-band-derived stellar mass by the model $K$-band stellar M/L.
Our assumptions are accurate to better than 40\% in both cases,
and make no difference to our results.

Each SDSS$+$2MASS galaxy is assigned a gas mass 
from a galaxy in the \citet{bdj} comparison sample 
with similar half-light radius and $K$-band
luminosity.  For galaxies morphologically classified as late-type we
assign gas masses using a randomly chosen comparison galaxy
that is within a factor of two in size, and within
one magnitude
in $K$-band luminosity.  Galaxies are morphologically classified
using the $r$-band concentration parameter following \citet{strat}.
We scale the gas mass by the difference
in luminosity to conserve the galaxy gas fraction.
As a consistency check, we also assign gas masses by
choosing the nearest galaxy in half-light radius--luminosity
space, and by assigning gas to galaxies of all morphological
types (see Fig. \ref{fig:mf}).  These changes make no appreciable difference
to our results.

\section{The Baryonic Galaxy Mass Function} \label{sec:results}

Our baryonic galaxy MF is 
shown as the solid line with open circles
in the left panel of Fig.\ \ref{fig:mf}.  
Shown for comparison is the stellar MF
from \citet{cole01}, and 
the $g$-selected stellar MF described above
and in \citet{bell03}.  The baryonic galaxy MF follows the stellar MF
at high masses, shows a modest offset at the 
`knee' of the MF, and shows a reasonably steep faint-end slope.  
This behavior is expected, as low-mass field galaxies tend to 
have high cold gas fractions and more ongoing SF
\citep{bdj,kauffmann03b}, which
steepens the baryonic MF compared to optical/NIR LFs that
typically have $\alpha \sim -1.0$.  The Schechter Function fit parameters
are $\phi^*\,h^{-3} = 0.0108(6)$\,Mpc$^{-3}$\,log$_{10}M^{-1}$, 
$M^*\,h^2 = 5.3(3)\times10^{10}\,M_{\sun}$, and $\alpha = -1.21(5)$,
where the formal error estimates are in parentheses.

In the middle panel of Fig.\ \ref{fig:mf},
we show the effects of the systematic uncertainties.  In particular,
the dashed, dotted and dot-dashed lines show the effects of 
changing the gas mass assignment (see the 
figure caption for more details); these uncertainties 
have only small effects.
The two bare solid lines show the effect of assuming different stellar IMFs.  
As stated earlier, we adopt as our default an 
IMF that has the largest stellar M/L permitted, without over-predicting
the rotation velocities of spiral galaxies in the Ursa
Major Cluster \citep[solid line with open circles;][]{ml}.  
Yet, the stellar M/L may be
lower than this maximal value; thus, we show two cases.  First, we plot (upper
solid line) the increasingly popular \citet{kroupa} and \citet{kennicutt} 
IMFs, which both have M/Ls of $\sim$70\% 
of maximal IMF ($\sim$50\% Salpeter).  The other
case is an IMF that has M/Ls of only 40\% of our maximal IMF (the lowest
solid line), which
corresponds roughly to the 63\% disk velocity 
contribution to the rotation curve 
as argued by \citet{bottema}.  
It is clear that, assuming a universal IMF, 
the factor-of-two uncertainty in the stellar M/L dominates the error
budget in terms of total cold baryonic mass in the local Universe.  Given
the stellar IMF, the systematic uncertainties 
are $\la 25$\%.  We estimate these uncertainties
using different passbands to estimate stellar mass,
using different gas mass assignment methods, and accounting 
for the effects of dust and bursts of SF on the M/Ls
\citep{ml,bell03}.
The Schechter Function fit parameters for the 
Kennicutt/Kroupa case and the Bottema case are:
$\phi^*\,h^{-3} = 0.0116(5),0.0142(8)$\,Mpc$^{-3}$\,log$_{10}M^{-1}$, 
$M^*\,h^2 = 3.78(11),2.24(8)\times10^{10}\,M_{\sun}$, and 
$\alpha = -1.22(3),-1.20(3)$.

It is important to make sure that our statistical procedure assigns gas
masses consistent with the true galaxy population by 
comparing with the \hi or \mol MF of galaxies (the right hand
panel of Fig.\ \ref{fig:mf}).
The solid black line with open circles and
error bars is our \hi MF derived in this way.  The dashed
and dotted lines show the effects of using the closest galaxy 
to estimate gas mass and allowing elliptical galaxies to have gas mass, 
respectively.  For comparison, the \hi MF of the blind \hi Aricebo survey
of \citet{rosenberg} is plotted as the smooth solid curve.  
We also show our prediction of the \mol galaxy MF as the lower solid grey line
with open circles and error bars.  For 
comparison, we plot the Schechter Fit to the first observational
estimate of the \mol MF \citep{keres03}.  Given the uncertainties, it is 
clear that our method for estimating galaxy gas masses reproduces
the \hi and \mol galaxy MFs with astonishing precision.  
Therefore, the sample of galaxies that 
we use to assign gas masses is indeed reasonably representative,
and lends considerable credibility to our estimate of the baryonic
galaxy MF.

\section{Discussion} \label{sec:disc}

Even with the factor-of-two uncertainty from the contribution 
of low-mass stars to the overall stellar M/L,
we can still draw some conclusions
about the local Universe.  It is clear that the 
overall efficiency of galaxy formation is very low.
Firstly, the faint end slope of the baryonic MF 
is $\sim -1.2$, which is much shallower than the $\sim -2$ expected
for the halo MF \citep[e.g.,][]{white91}.  Secondly, integrating
under the MF, we derive $\Omega_{\rm cold\,\,baryon}\,h 
= 2.4^{+ 0.7}_{- 1.4}\times10^{-3}$, including the
IMF and 25\% systematic stellar M/L uncertainties.  Our estimate agrees well
with the value of $2.9\pm1.5\times10^{-3}$ from \citet{fuku}, and is
preferred due to our better accounting for stellar M/Ls compared with
\citet{fuku} who use (harder to convert into stellar mass) $B$-band luminosity
densities assuming a similar IMF to the maximum-disk IMF we adopt here.
Taking the value of the total baryon density from the Big Bang
nucleosynthesis value of \citet{omeara},
and assuming h=0.7$\pm$0.07, we find 
$\Omega_{\rm cold\,\,baryon}/\Omega_{\rm b}
\sim 8^{+4}_{-5}$\%, where the error estimates account for the 
uncertainties in IMF, H$_0$, $\Omega_{\rm b}$, our gas assignment method, and
the $\la$25\% uncertainties in stellar M/Ls from 
dust and bursts of SF.  Our value is quite consistent
with the low galaxy formation efficiency characteristic of 
most current models, which 
have low efficiencies at the low and high-mass ends because of
feedback from supernovae and  inefficient gas 
cooling, respectively \citep[e.g.,][]{cole00}.

Accounting for the possible gaseous content of elliptical galaxies,
for sub-maximal M/Ls, and for the effects of dust and bursts of SF
on stellar M/Ls, the universal gas fraction, $f_g = \Omega_{\rm cold\,\,gas}/\Omega_{\rm cold\,\,baryons}$, should lie in the range $0.2 \la f_g \la 0.5$.
For the `maximal' IMF, we find $f_g\sim 25$\%.
\citet{fuku} and \citet{keres03} find values of 15--20\%
when our IMF is adopted; their slightly lower determinations stem
primarily from a lower estimate of \hi mass 
density.  Nevertheless, all the studies agree that $f_g \leq 0.5$; therefore,
the dynamically cold baryons (i.e.,
the gas and stars in disks and spheroids) are primarily
in the form of stars, even for low stellar M/Ls. 

It is well-known observationally that cluster optical/NIR
LFs have steeper faint-end slopes than field LFs \citep[e.g.,][]{trentham02}.
Furthermore, cluster galaxies tend to have little ongoing 
SF and little gas, so that most cluster
galaxies are star-dominated with large stellar M/Ls
\citep[e.g.,][]{kuntschner}.
Thus, the trend of increasing faint end slope with increasing 
cluster mass noted by e.g., \citet{trentham02} may be 
more naturally interpreted as a constant baryonic
MF, with a suppression of recent
SF in massive clusters of galaxies.
Obviously, a deeper investigation of this issue is warranted before
speculating any further.

\section{Conclusions} \label{sec:conc}

Together with the baryonic (stellar$+$cold gas) luminosity--linewidth
relation \citep[e.g.,][]{mcgaugh00,ml}, the baryonic galaxy
MF is an ideal test of models of galaxy formation and evolution.
In this {\it Letter}, we have estimated the baryonic galaxy MF
in the local Universe for the first time assuming a
universally-applicable stellar IMF.
We assign gas fractions statistically to a large
sample of galaxies from 2MASS and SDSS, using a local
sample with accurate $K$-band and gas fraction data.
We cross-check this statistical procedure
against independent \hi and \mol surveys, finding excellent agreement.
The baryonic MF is similar to the 
stellar MF at the high mass end (with a slightly 
higher density normalization), and has a reasonably
steep faint end slope, $\alpha \sim -1.2$, due to the 
typically high cold gas fractions and low stellar M/Ls of low-mass
galaxies.  Integrating under the 
baryonic MF, we find that the 
baryonic (stellar$+$cold gas) mass density implied by this
estimate is $\Omega_{\rm cold\,\,baryon} = 
2.4^{+ 0.7}_{- 1.4}\times10^{-3}$, or 8$^{+4}_{-5}$\% of 
the Big Bang nucleosynthesis
expectation.  This clearly implies a low overall efficiency of 
galaxy formation.  

\acknowledgements

E.\ F.\ B.\ was supported by the European Research Training
Network {\it Spectroscopic and Imaging Surveys for Cosmology}.
D.\ H.\ M.\ and M.\ D.\ W.\ acknowledge support
by JPL/NASA through the 2MASS core science projects.
This publication makes use of data products from the 
{\it Two Micron All Sky Survey} 
which is a joint project of the 
University of Massachusetts and the Infrared Processing and 
Analysis Center/California Institute of Technology, funded by 
the National Aeronautics and Space Administration and 
the National Science Foundation.
This publication also makes use of the 
{\it Sloan Digital Sky Survey} (\texttt{http://www.sdss.org/}).  
We thank the referee, Andrea Ferrara, for his
helpful suggestions and feedback.

\end{document}